\documentclass[vecphys]{svmult}

\usepackage{makeidx}         % allows index generation
\usepackage{graphicx}        % standard LaTeX graphics tool
\usepackage{multicol}        % used for the two-column index
\usepackage[bottom]{footmisc}% places footnotes at page bottom
\usepackage{amssymb}
\usepackage{cite}
\makeindex             % used for the subject index
\begin{document}

\title*{Modeling Metropolis Public Transport }
\author{Christian von Ferber\inst{1}$^,$\inst{2}, Taras Holovatch\inst{3},
Yurij Holovatch\inst{4}$^,$\inst{5}\and Vasyl Palchykov\inst{4}}
\authorrunning{C. von Ferber, T. Holovatch, Yu. Holovatch, V. Palchykov}
\institute{Applied Mathematics Research Centre, Coventry University,
Coventry CV1 5FB, UK \texttt{C.vonFerber@coventry.ac.uk}
 \and
 Physikalisches Institut, Universit\"at Freiburg, 79104 Freiburg,
Germany
 \and
 Ivan Franko National University of Lviv, 79005 Lviv,
Ukraine
 \and
 Institute for Condensed Matter Physics of the National Academy of Sciences of Ukraine,
 79011 Lviv, Ukraine
 \and
 Institut f\"ur Theoretische Physik, Johannes Kepler Universit\"at Linz, 4040 Linz, Austria
 \texttt{hol@icmp.lviv.ua}}

\maketitle

\abstract{We present results of a survey of public transport
networks (PTNs) of selected 14 major cities of the world with PTN
sizes ranging between 2000 and 46000 stations and develop an
evolutionary model of these networks. The structure of these PTNs is
revealed in terms of a set of neighbourhood relations both for the
routes and the stations. The networks defined in this way display
distinguishing properties due to the constraints of the embedding 2D
geographical space and the structure of the cities. In addition to
the standard characteristics of complex networks like the number of
nearest neighbours, mean path length, and clustering we observe
features specific to PTNs. While other networks with real-world
links like cables or neurons embedded in two or three dimensions
often show similar behavior, these can be studied in detail in our
present case. Geographical data for the routes reveal surprising
self-avoiding walk properties that we relate to the optimization of
surface coverage. We propose and simulate an evolutionary growth
model based on effectively interacting self-avoiding walks that
reproduces the key features of PTN.}

\section{Introduction}
\label{sec:1}
% Always give a unique label
% and use \ref{<label>} for cross-references
% and \cite{<label>} for bibliographic references
% use \sectionmark{}
% to alter or adjust the section heading in the running head

Urban public transport networks (PTNs) share general features of other
transportation networks like airport, railroad networks, power grids,
etc.  \cite{Newman03}. The most obvious common features are their
embedding into a 2D space, evolutionary growth, and optimization.  The
evolution of specific local transportation networks is closely related
to that of the town or region in which they are embedded. However,
statistically, some general overall features such as their fractal
dimensions have been observed \cite{Benguigui}.  Modeling different
aspects of transportation network structure and functioning helps to
understand and optimize various processes that occur on these networks
as well as to improve their planning. In this paper, we will consider
a PTN from the point of view of complex network theory
\cite{Newman03,Albert02,Dorogovtsev03,Holovatch06} quantifying
statistical features of their structure and proposing a model that
reproduces their key features. To exemplify these, we survey the
empirical analysis of PTNs of 14 major cities of the world (see
Refs. \cite{Ferber07a,Ferber07b} for more details). While the
empirical analysis of PTNs of different cities has been subject of
several studies
\cite{Marchiori00,Latora01,Latora02,Seaton04,Ferber05,Sienkiewicz05a,%
Ferber07a,Xu07,Holovatch07} (see Sec. \ref{sec:2}), we are not aware
of previous attempts to specifically model the network evolution of a PTN
in the way we propose here.
The set-up of the paper is as follows: in the next section we
discuss common features of PTNs and list some numbers that quantify
these features; section \ref{sec:3} we describe our model
for PTN evolution and give supporting arguments; in section
\ref{sec:4} we compare some characteristics of real and simulated
PTNs; conclusions and an outlook are given in section \ref{sec:5}.

\section{Empirical analysis of PTNs: statistical properties}
\label{sec:2}

A distinct feature of our study is that we interpret the PTN as a
network of {\em all} means of public transport (buses, trams, subway,
etc.) offered in a given city.  Several previous studies have analyzed
specific sub-networks of PTNs. The Boston
\cite{Marchiori00,Latora01,Latora02,Seaton04} an Vienna
\cite{Seaton04} subway networks may serve as examples. However, each
particular sub-network (e.g. the network of buses, trams, or
subways) is not a closed system: it is a subgraph of a wider
transportation system of a city, or as we call it here, of a PTN.
Therefore to understand and describe the properties of transport in a
city it is important to deal with the complete PTN, not restricting the analysis
to specific parts. Indeed, extending the restricted subway network
to the ``subway + bus" network drastically changes the network
properties, as it was shown for Boston \cite{Latora01,Latora02}.

Another important quantity that certainly restricts the
reliability of grneralizations drawn from trends observed for specific PTNs
of different cities is the statistics, i.e. the size of the observed PTNs.
The numbers of stations $N$ in the PTNs analyzed so far ranged from
several decades (subway networks of Boston, $N=124$ \cite{Latora02},
and Vienna, $N=76$ \cite{Seaton04}) to several thousands as in the
PTN analysis of 22 Polish cities with up to 2811 stations
\cite{Sienkiewicz05a} or bus-transport networks of three Chinese
cities with up to $N=3938$ stations. In our sampling
\cite{Ferber05,Ferber07a} we have chosen PTNs of 14 major cities of
the world of various size with $N$ ranging from 1544 (D\"usseldorf)
to 46244 (Los Angeles), see Table \ref{tab:1}.

\begin{figure}
\centering
\includegraphics[width=70mm]{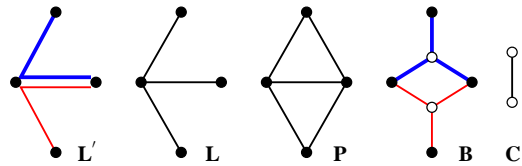}
\caption{Different representations of a PTN.
$\mathbb{L}^\prime$-space: nodes correspond to stations, links show
different routes. $\mathbb{L}$-space: the same as
$\mathbb{L}^\prime$ but multiple links are reduced to single ones.
$\mathbb{P}$-space: any two stations (nodes) are connected when they
are serviced by a common route. $\mathbb{B}$-space: routes (open
discs) are linked to those stations (filled discs) that they service.
$\mathbb{C}$-space: nodes correspond to routes, links show that two
routes share a common station. Note that  by  keeping multiple links
in $\mathbb{P}$ and $\mathbb{C}$ one may also define $\mathbb{P}^\prime$
and $\mathbb{C}^\prime$-spaces.}
 \label{fig:1}
\end{figure}
Here, we define different graph representations (`spaces') for a given PTN
in terms of nodes (vertices) and links (edges) as illustrated in Fig. \ref{fig:1}.
The primary network topology is defined by a set of routes each
servicing an ordered series of given stations, see the sketch labeled
$\mathbb{L}^\prime$ in Fig. \ref{fig:1}
Based on this we define a graph ($\mathbb{L}$-space) representing each station
by a node and linking any two that are served consecutively by at least
one route. More generally linking any two stations serviced by a common route
we define the $\mathbb{P}$-space \cite{Sen03,Sienkiewicz05a} representation.
The interrelation of the routes in turn is described by a complementary
$\mathbb{C}$-space representation where now the nodes represent routes
and linking any two that service a common station.
From Fig. \ref{fig:1} it is easy to verify, that the last two
representations correspond to one-mode projections of a bipartite
$\mathbb{B}$-space represenation with nodes of two types, representing
either a station or a route linking each route with all stations it services.

\begin{table}
\centering \caption{PTN characteristics in $\mathbb{L}$- and
$\mathbb{P}$- representations, see text for definitions.
$N$: number of stations, $M$: number
of routes, the scale $\hat{k}$ or the exponent $\gamma$ of fits
to the node degree distribution $p(k)$ to the laws (\ref{eq:1}) or
(\ref{eq:2}) respectively, $\mathcal{C}$: ratio of the mean
clustering coefficient to its random graph value, maximal $\hat{\ell}$,
and mean $\overline{\ell}$ shortest path length.
Characteristics in $\mathbb{P}$-space representations are indicated by a
subscript `p'. See Refs. \cite{Ferber07a,Ferber07b} for more data and
sources. \label{tab:1}} \tabcolsep=1mm
\begin{tabular}{lllllllllll}
 \hline\noalign{\smallskip} City &  $N$ & $M$ &
& & $\mathcal{C}$ & $\mathcal{C}_{\rm p}$ & $\hat{\ell}$ &
$\overline{\ell}$ &
 $\hat{\ell}_{\rm p}$ &  $\overline{\ell_{\rm p}}$ \\
 \noalign{\smallskip}\hline\noalign{\smallskip}
Berlin &  2996 & 218 &    $\hat{k}$=1.24 & $\hat{k}_{\rm p}$=38.5 & 52.85 & 42.03 & 68 & 18.61 & 5 &  2.93 \\
Dallas &  6571 & 131 &   $\gamma$=4.99  &   $\hat{k}_{\rm p}$=76.9 &  17.26 &  63.00 & 269 & 85.84 & 10 &  3.78 \\
D\"usseldorf &  1544 &   124  & $\hat{k}$=1.12 &  $\hat{k}_{\rm p}$=58.8 & 22.45 & 20.97 & 56 & 13.18 & 5 &  2.58 \\
Hamburg &  8158 & 708 &  $\hat{k}$=1.47 &  $\hat{k}_{\rm p}$=55.6 &   262.92 &  133.99 & 158 & 39.74 & 11 &  4.79 \\
Hong Kong &  2117 & 321 &  $\hat{k}$=2.60 &  $\hat{k}_{\rm p}$=125.0 &  58.98 &  12.51 & 60 & 11.11 & 4 &  2.26 \\
Istanbul &  4043 & 414 &   $\gamma$=4.04  &  $\hat{k}_{\rm p}$=71.4 &   41.40 &   41.54 & 131 & 29.69 & 6 &  3.09 \\
London &  11012 & 2005 &   $\gamma$=4.58  &  $\gamma_{\rm p}$=4.39 &   326.17 &   90.00 & 107 & 26.68 & 6 &  3.26 \\
Los Angeles &  46244 & 1893 & $\gamma$=4.88 &  $\gamma_{\rm p}$=3.92  &  588.44 &  427.06 & 247 & 43.55 & 14 &  4.60 \\
Moscow &  3755 & 679 &   $\hat{k}$=2.12 &   $\hat{k}_{\rm p}$=50.0 &   128.23 &   41.93 & 28 &  7.08 & 5 &  2.52 \\
Paris &  4003 & 232 &   $\gamma$=2.61  &   $\gamma_{\rm p}$=3.70  &   85.90 &   71.75 & 47 &  7.22 & 5 &  2.79 \\
Rome &  6315 & 681 &   $\gamma$=4.39 &   $\hat{k}_{\rm p}$=45.5 &  68.61 &   76.93 & 93 & 29.64 & 8 &  3.58 \\
S\~ao Paolo  & 7223 & 998 & $\gamma$=2.72 & $\hat{k}_{\rm p}$=200.0  & 268.83  & 38.32 & 33 & 10.34 & 5 &  2.66 \\
Sydney  & 2034 & 596 &   $\gamma$=3.99 &   $\hat{k}_{\rm p}$=38.5 &  81.62 &  34.92 & 35 & 12.76 & 7 &  3.03 \\
Taipei  & 5311 & 389 &  $\hat{k}$=1.75 &   $\hat{k}_{\rm p}$=200.0 &  186.23 &  15.38 & 74 & 20.86 & 6 &  2.35 \\
\noalign{\smallskip}\hline
\end{tabular}
\end{table}

In our analysis, we are interested in different features of the
PTN as measured when represented in the above defined spaces.
It is worth to mention here, that these standard network characteristics
measured in different spaces turn out to be specific
of practical value in judging about the quality of public
transport in a given city. The particular quantities we analyze here
are the maximal and mean shortest path length $\hat{\ell}$ and
$\overline{\ell}$, the clustering coefficient $C$, and the betweenness $C_B$;
for definitions see Ref. \cite{attack} in this volume.
Table \ref{tab:1} gives some of these quantities for the cities
analyzed in $\mathbb{L}$- and $\mathbb{P}$- representations. One can
see that the above networks are highly clustered small worlds
characterized by small
shortest path lengths and large ratios $\mathcal{C}$ of the mean
clustering coefficient relative to its value $C_{ER}=2M/N^2 $
on a random graph with the same numbers of nodes $N$ and links $M$.

\begin{figure}
\centering
\includegraphics[width=55mm]{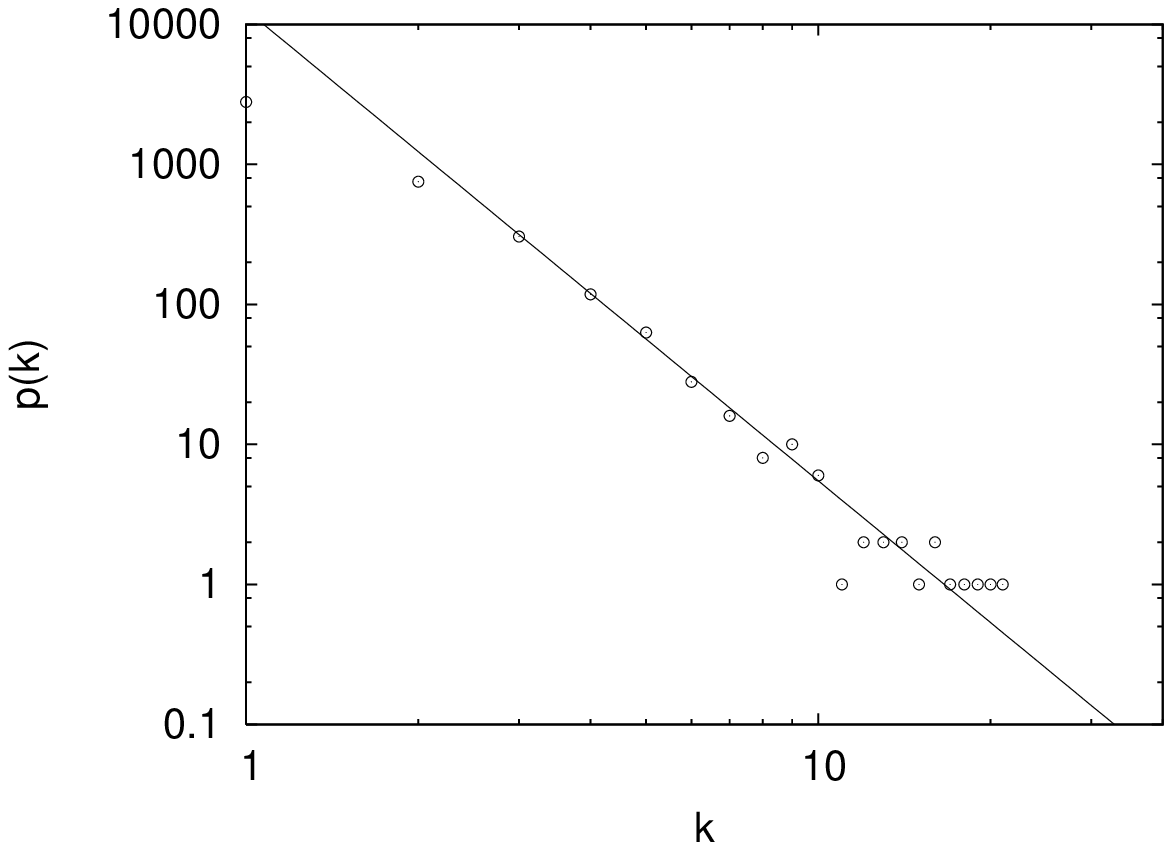}
\includegraphics[width=55mm]{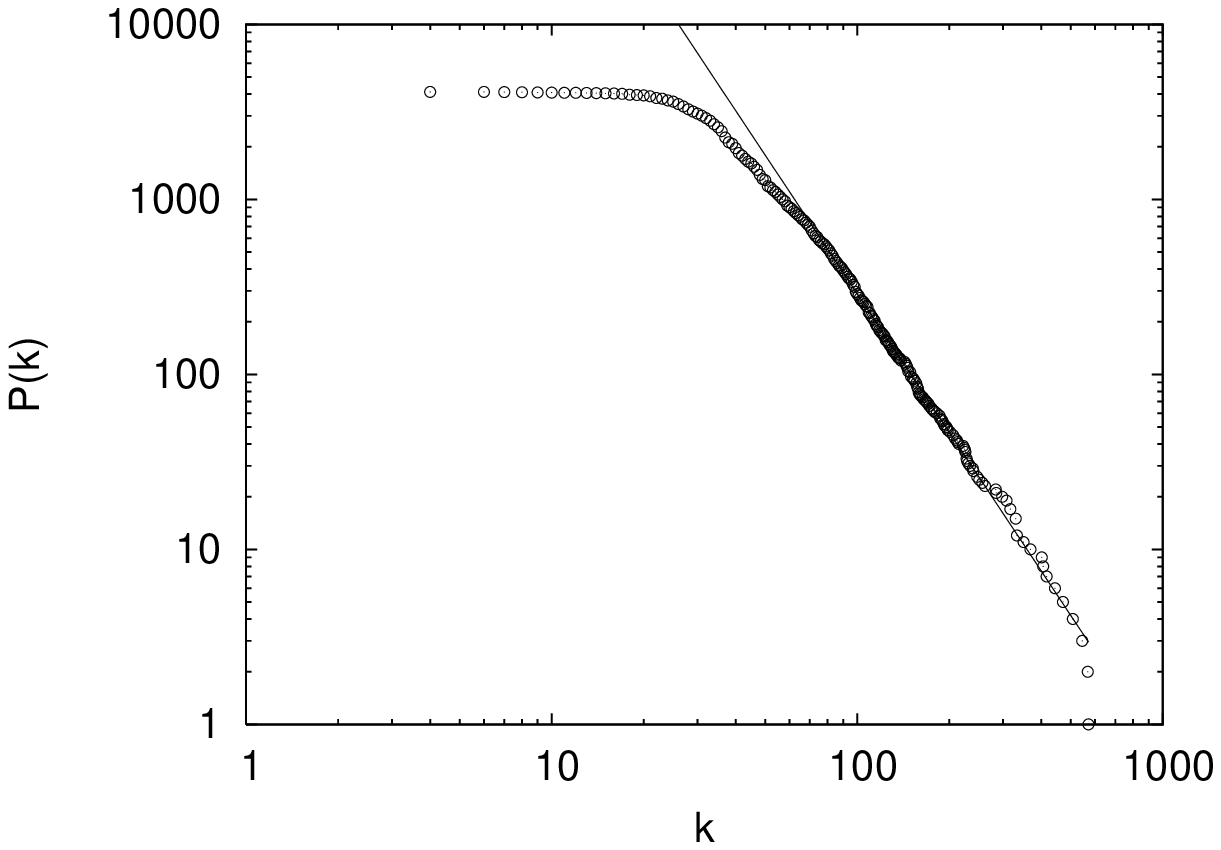}
\includegraphics[width=55mm]{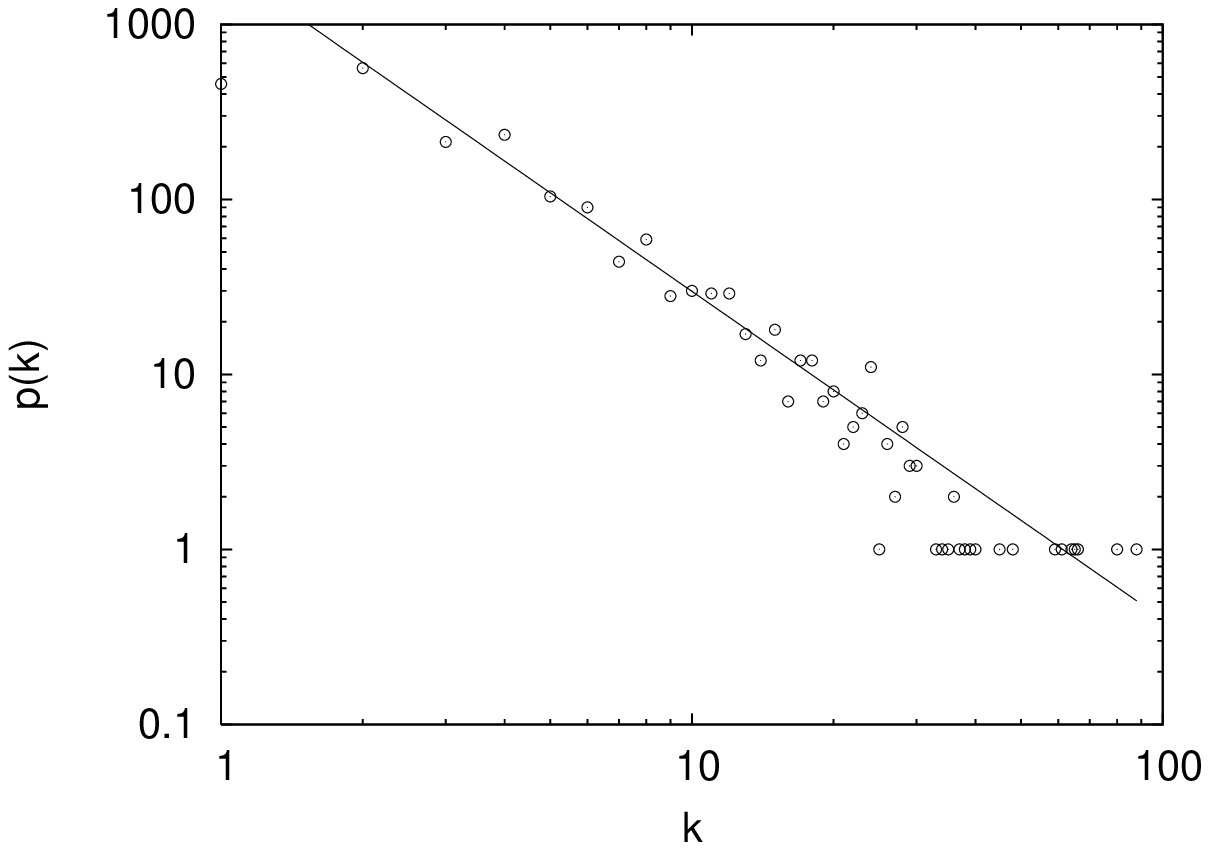}
\includegraphics[width=55mm]{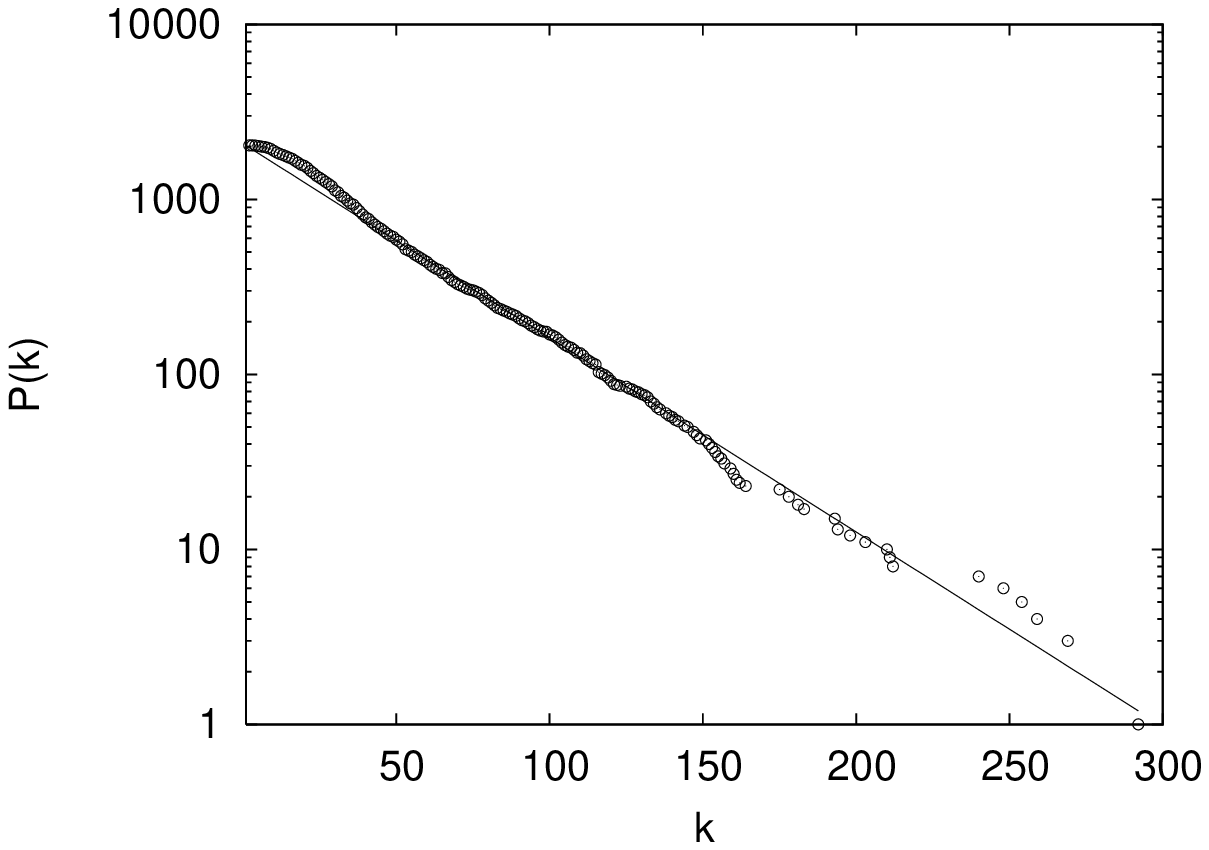}
\caption{Node degree distributions for Paris (first row) and Sydney
(second row) PTNs. Left column: degree distributions $p(k)$ in
$\mathbb{L}$-space. Right column: cumulative degree distributions
(\ref{eq:3}) in $\mathbb{P}$-space.}
 \label{fig:2}
\end{figure}

To classify the node degree distributions $p(k)$ we performed fits to
both an exponential function
\begin{equation}
\label{eq:1}
 p(k)=Ae^{-k/\hat{k}},
\end{equation}
as well as to a power law:
\begin{equation}
\label{eq:2}
 p(k)=Bk^{-\gamma}.
\end{equation}
The result of the better fit together with the value of the fit
parameters $\hat{k}$ or $\gamma$ are shown in table \ref{tab:1}.
One can see that the $\mathbb{L}$-space node degree distribution
of a part of these PTNs (8 out of 14) is governed by
power laws, indicating
scale-free properties. For the PTNs of several cities, this fact has
also been described in Refs. \cite{Ferber05,Sienkiewicz05a}. The
remarkable feature of the data shown in table \ref{tab:1} is that some
PTNs (3 out of 14) show scale-free behavior even in $\mathbb{P}$-space.
As examples compare the degree distributions $p(k)$ of the PTNs of
Paris and Sydney in Fig. \ref{fig:2}. Whereas the Paris PTN is scale-free both in
$\mathbb{L}$- and $\mathbb{P}$-space, the PTN of Sydney is scale-free in
$\mathbb{L}$-space only. Note that to reduce the noise in the data
we plot the $\mathbb{P}$-space cumulative degree distribution
\begin{equation}\label{eq:3}
P(k)=\sum_{q=k}^{\infty} p(q) .
\end{equation}

\begin{figure}
\centering
\includegraphics[width=55mm]{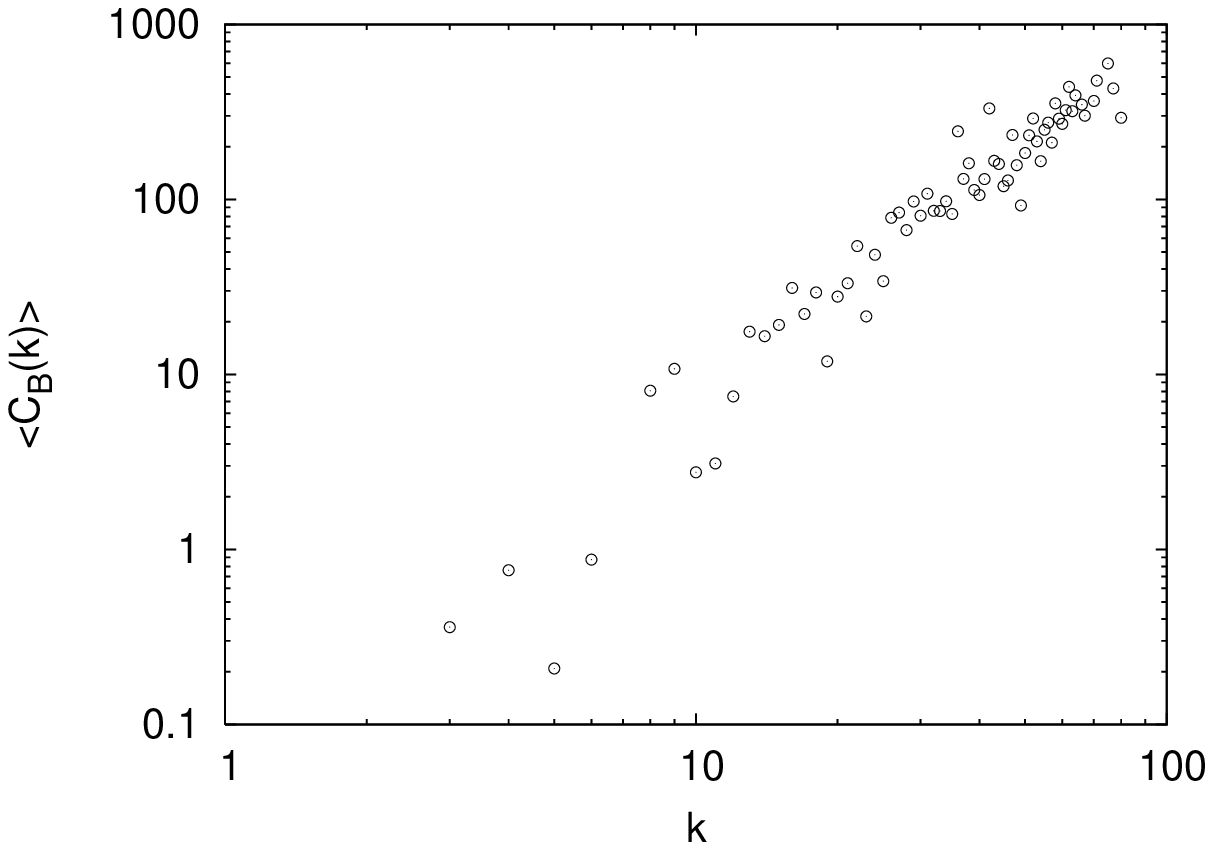}
\includegraphics[width=55mm]{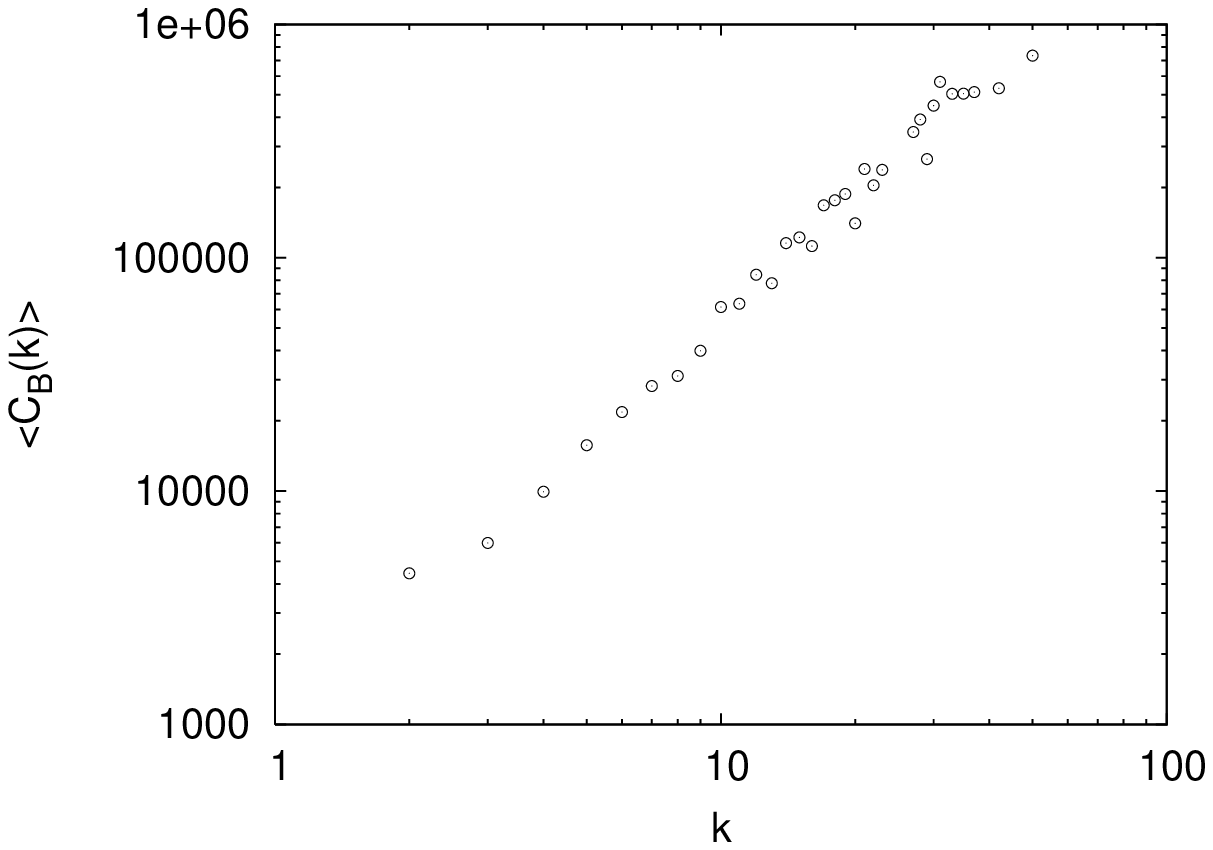}
\includegraphics[width=55mm]{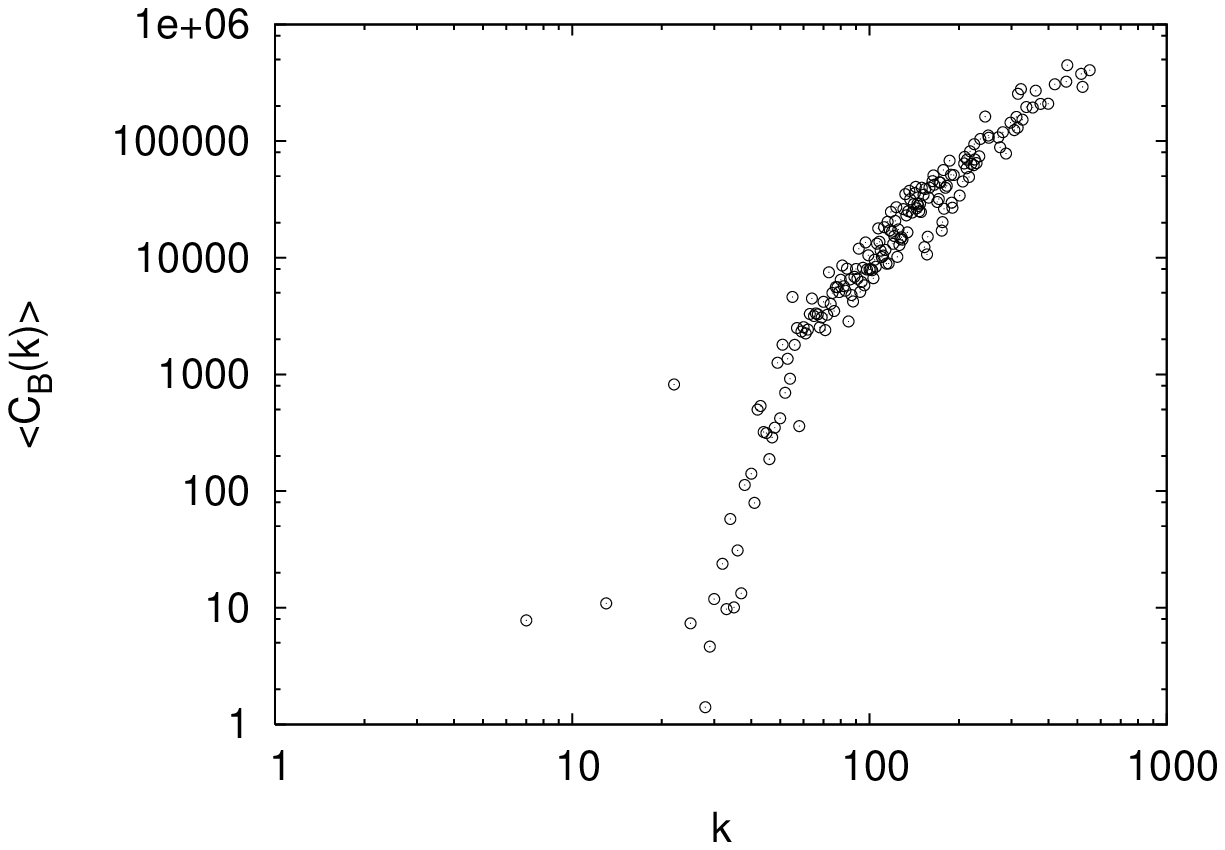}
\includegraphics[width=55mm]{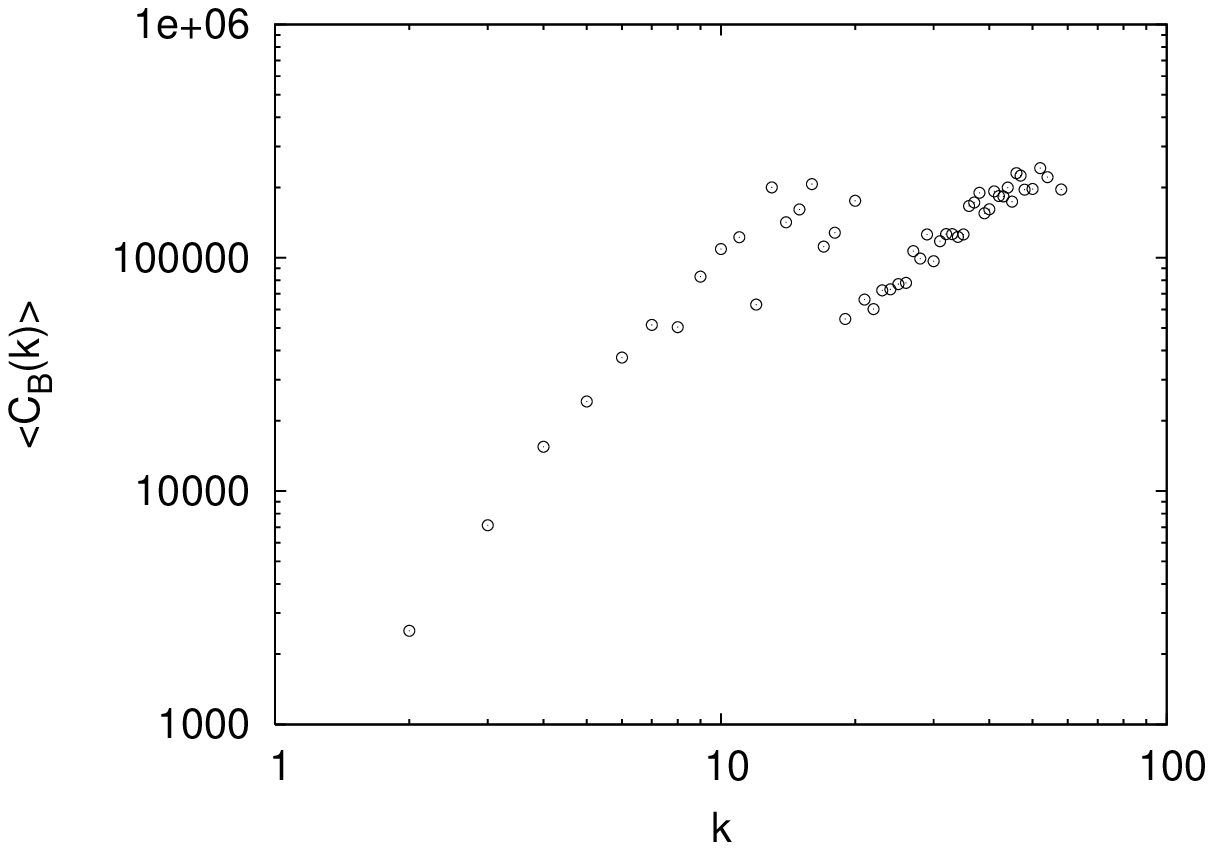}
\caption{Betweenness-degree correlations for Paris PTN in
$\mathbb{C}$-, $\mathbb{L}$-, $\mathbb{P}$-, and
$\mathbb{B}$-spaces.}
 \label{fig:3}
\end{figure}

In Fig. \ref{fig:3} plot the mean betweenness value $C_B(k)$ (see
\cite{attack} in this volume) as function of the node degree $k$ in
in $\mathbb{C}$-, $\mathbb{L}$-, $\mathbb{P}$-, and
$\mathbb{B}$-spaces (similar behaviour is found
 for the networks of other cities). One definitely sees a pronounced
correlation and a tendency for power-law behaviour: a phenomenon observed previously
for several other networks \cite{Sienkiewicz05a,Vazquez02,Goh03}.

\begin{figure}
\centering
\includegraphics[width=37mm]{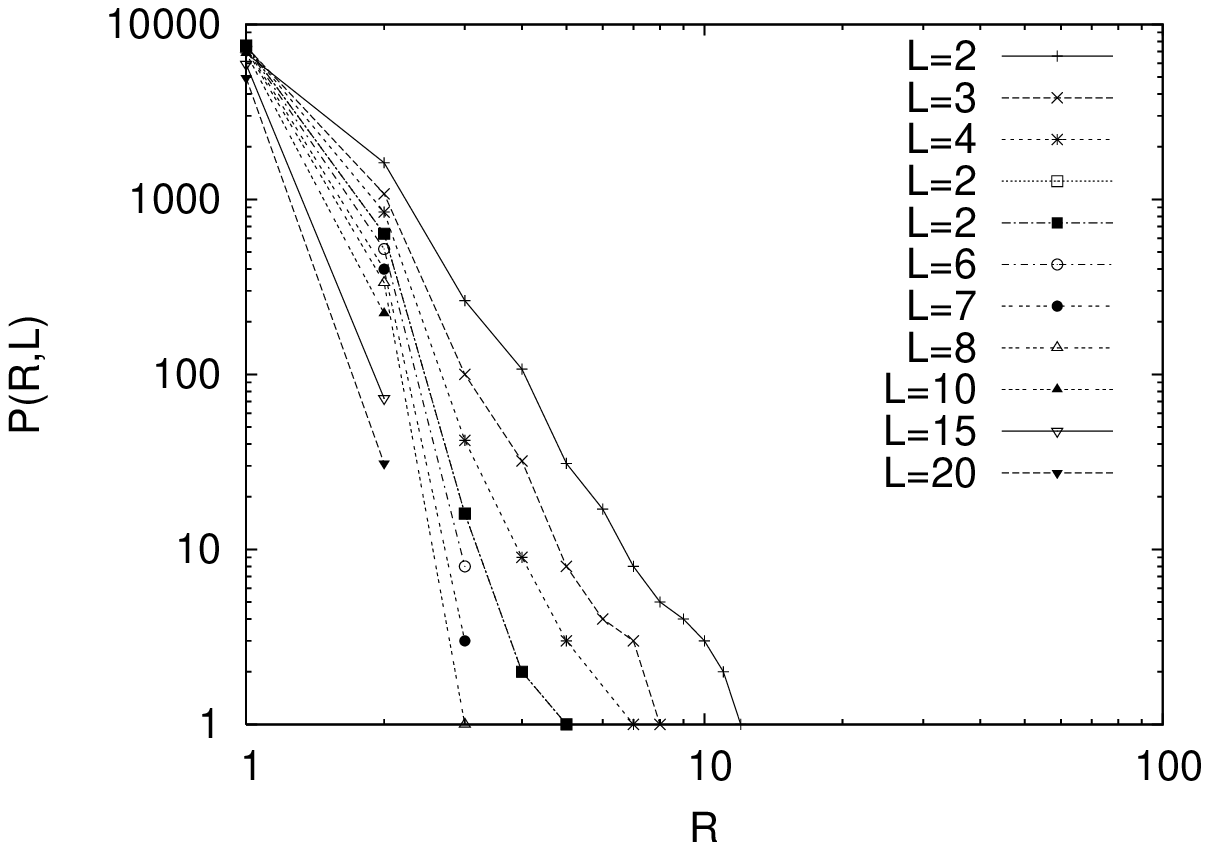}
\includegraphics[width=37mm]{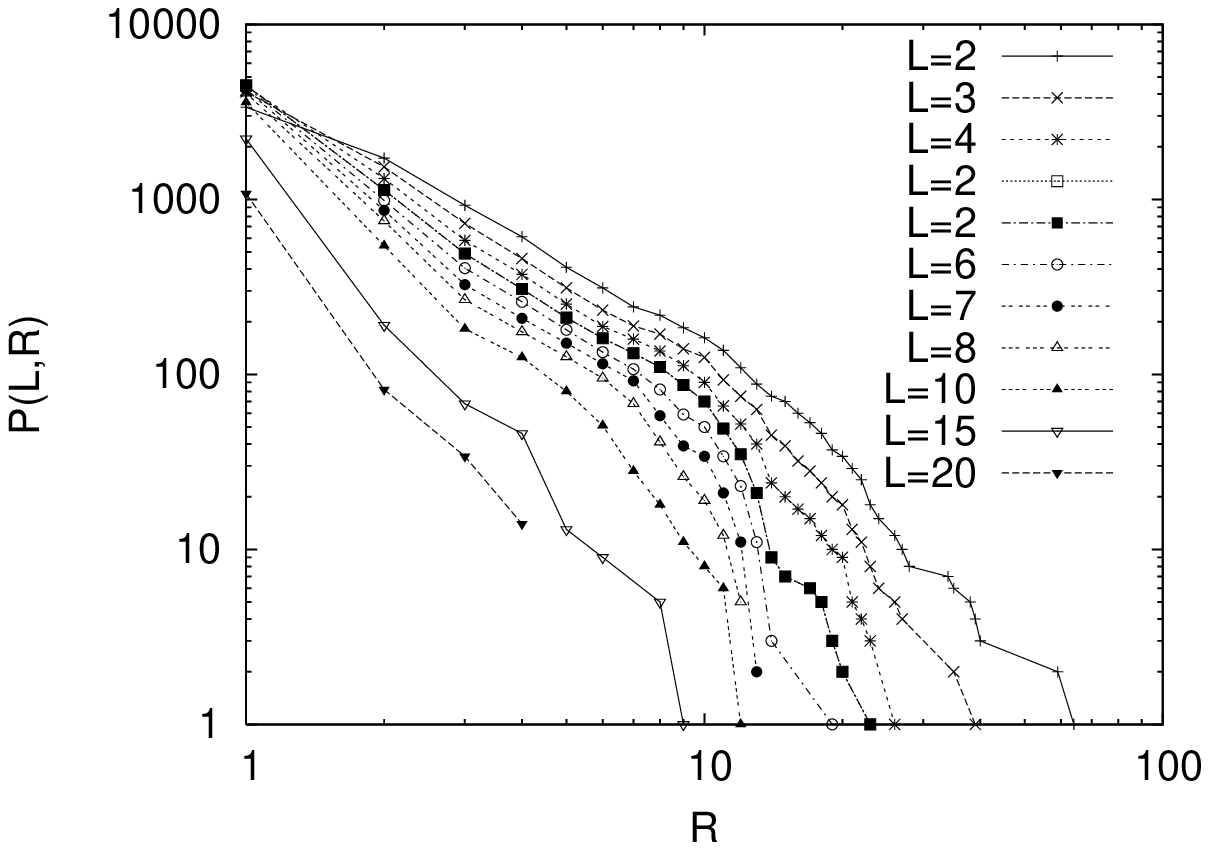}
\includegraphics[width=37mm]{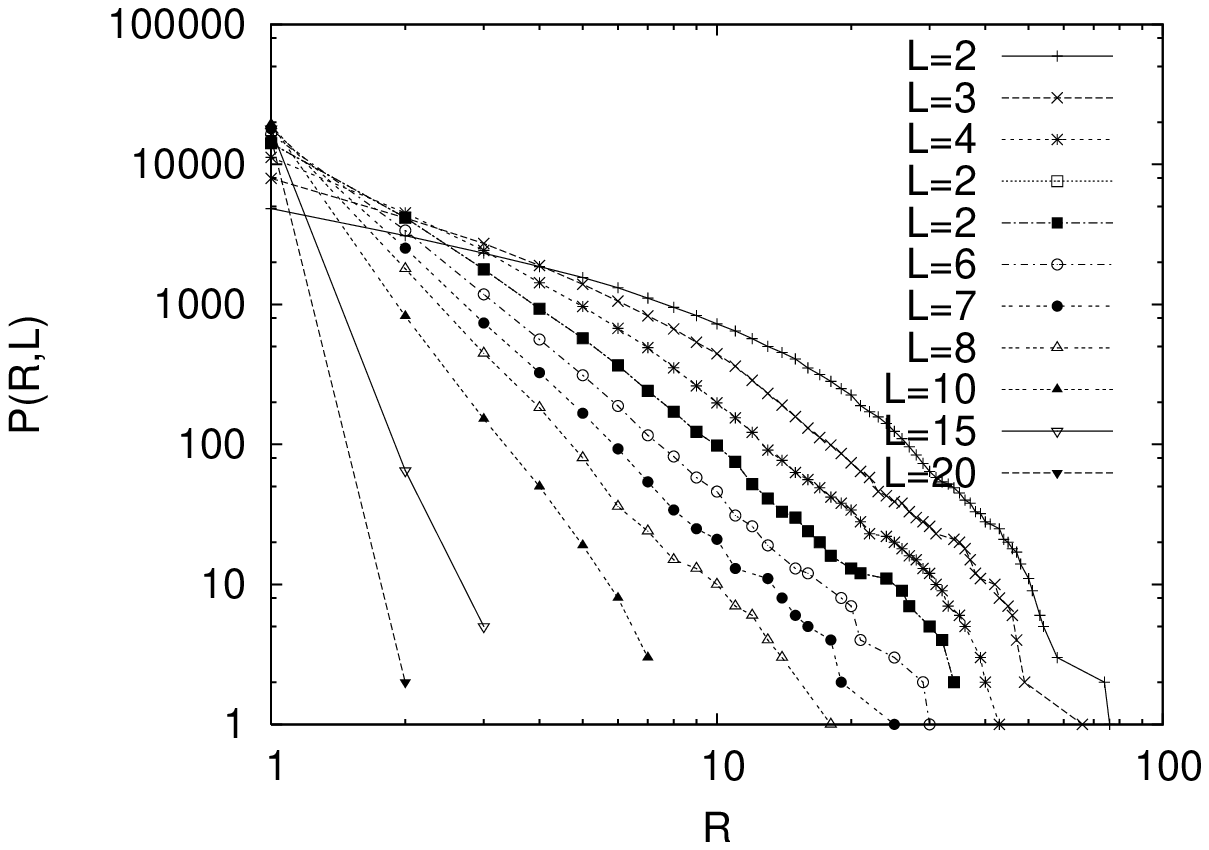}

{\bf{a)} \hspace{11em} \bf{b)} \hspace{11em} \bf{c)}}
 \caption{Harness distribution for Paris ({\bf a}), for Sydney ({\bf b}), and for
a simulated city ({\bf c}).}
 \label{fig:4}
\end{figure}

In addition to the standard characteristics of complex networks
discussed above, one can introduce some characteristics which are
specific to PTNs. One of them is due to the fact that very often
several routes go in parallel and pass $L$ consecutive stations. In
particular, the notion of {\em harness} was introduced in Ref.
\cite{Ferber07a} to quantify this behaviour. In Fig. \ref{fig:4} we
show the harness distribution $P(L,R)$: the number of sequences of
$L$ consecutive stations that are serviced by $R$ parallel routes
for the PTNs of Paris (a) and of Sydney (b). The log-log plot
indicates scale free properties of this distribution.
We have found similar behaviour for the majority of the cities
under consideration.

Following  this analysis of PTNs of different
cities of the world and having at hand the numerical data on
different features of these networks, let us propose
a model, that may reproduce most of these features. In particular, the model
should be capable to discriminate between different types of
behaviour observed so far in PTNs (an example may be the exponential
and power-law node degree distributions for $p(k)$ observed for
different cities).

\section{Evolutionary Model of PTNs}
\label{sec:3}

To model the properties of PTNs we have proposed \cite{Ferber07a} an
evolutionary growth model for these networks along the following
lines: We model the grid of streets by a square lattice and allow
every lattice site $\vec{r}$ (street corner) to be a potential
station visited by say $k(\vec{r})$ routes. The routes are modeled
as self-avoiding walks (SAWs) on this lattice. The rules of our
model are the following:
\begin{itemize}
 \item [1.] First route: construct a SAW of length $n$ starting at an arbitrary site.
 \item [2.] Subsequent routes:
 \begin{itemize}
 \item (i) choose a terminal station on
 lattice site $\vec{r}$ with probability $q\propto k(\vec{r}) +
 a$;
 \item (ii) choose a subsequent station of this route at a neighboring
 site $\vec{r}'$ with probability $q\propto k(\vec{r}') + b$;
 \item (iii) repeat step (ii) until the walk has reached $n$  stations,
 in case of self-intersection discard the walk and restart with
 step (i).
 \end{itemize}
 \item [3.] Repeat step 2 until $M$ routes are created.
\end{itemize}

\begin{figure}
\centering
\includegraphics[width=55mm]{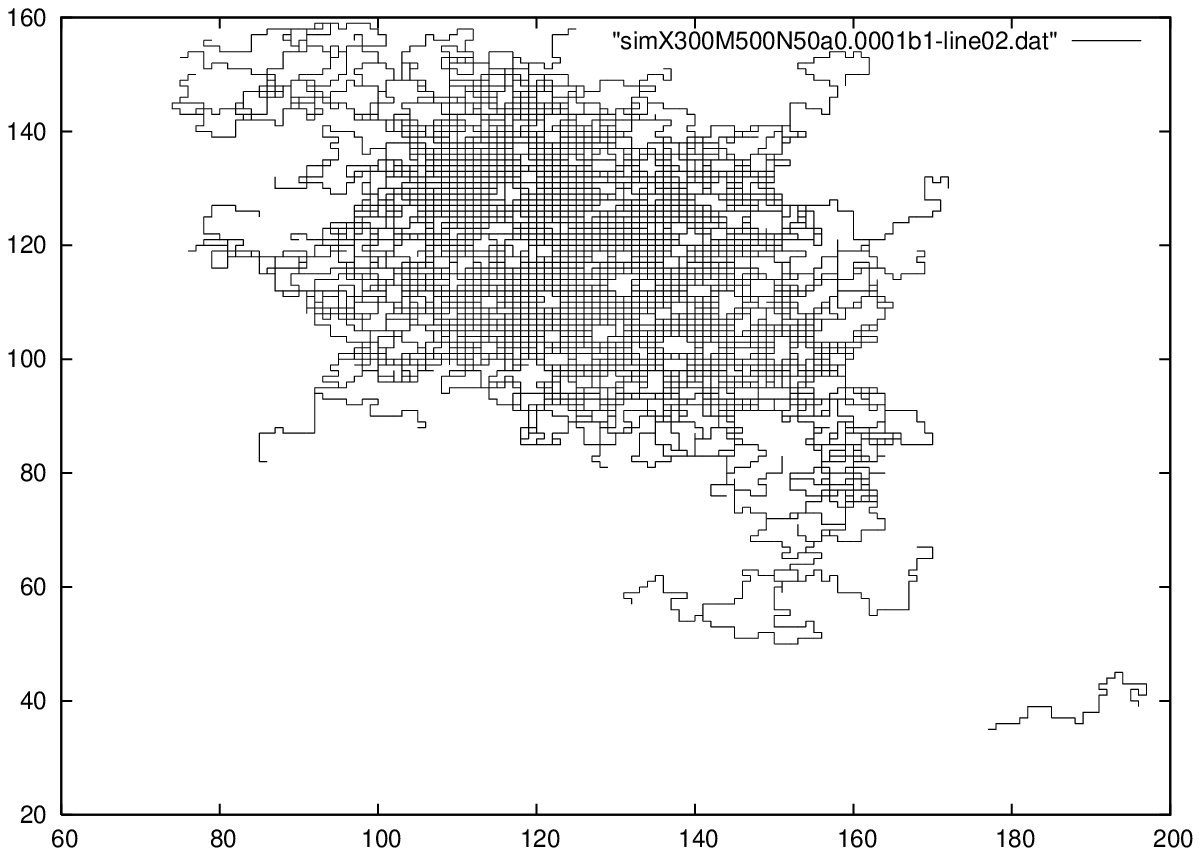}
\includegraphics[width=55mm]{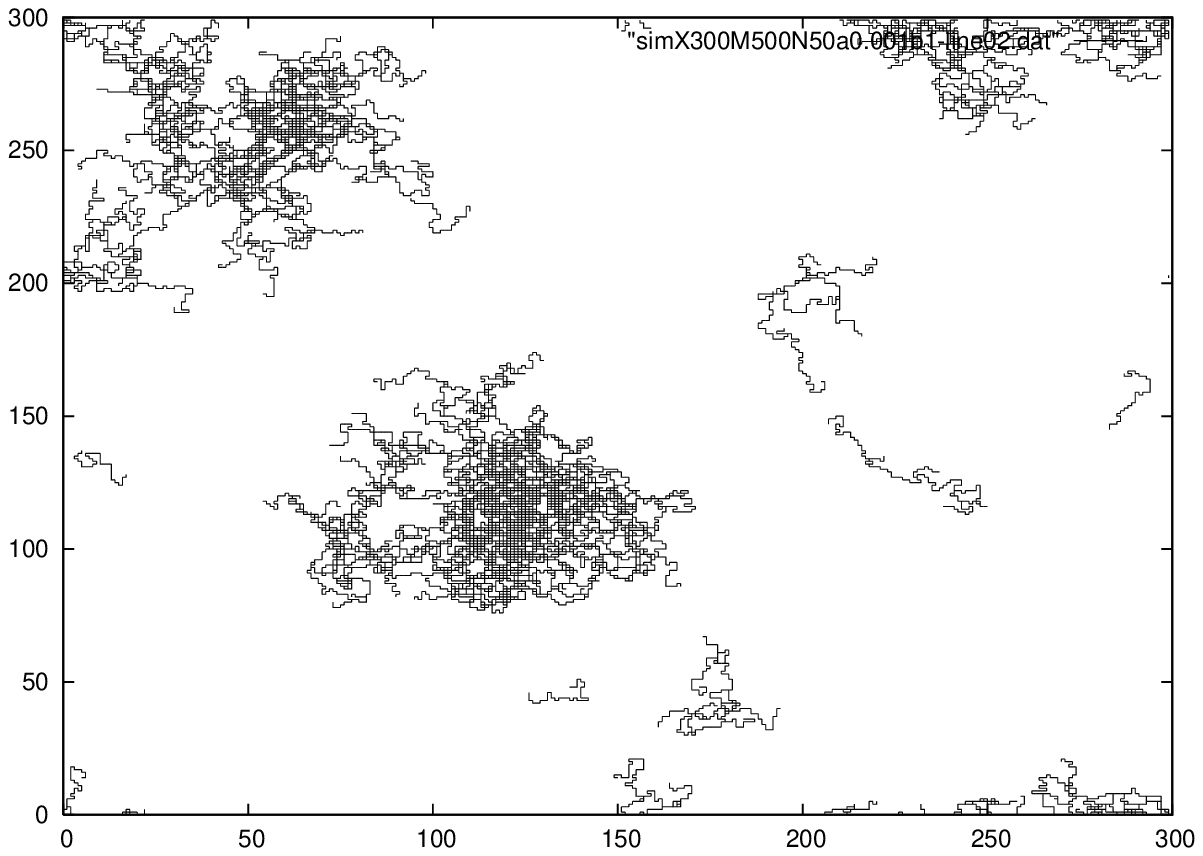}

\centerline{\bf{a)} \hspace{11em} \bf{b)} \hspace{11em}}
\includegraphics[width=55mm]{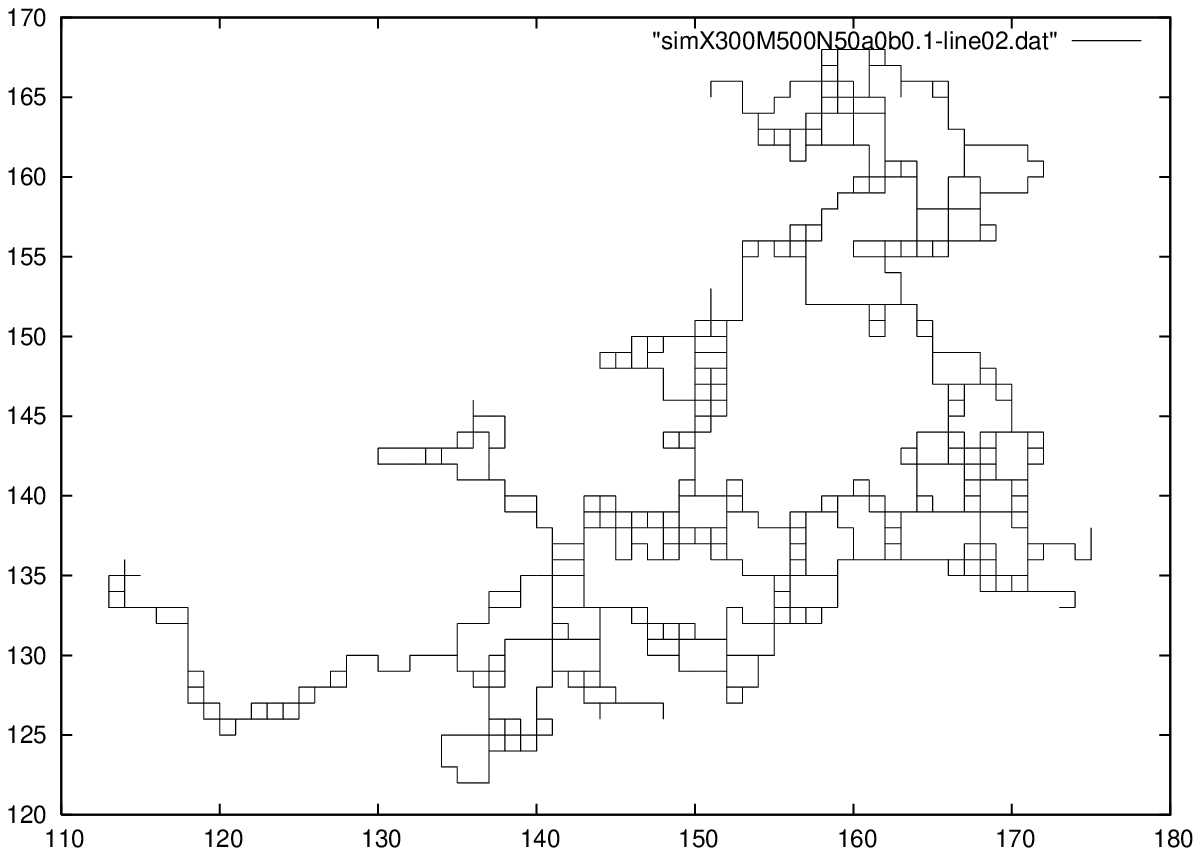}
\includegraphics[width=55mm]{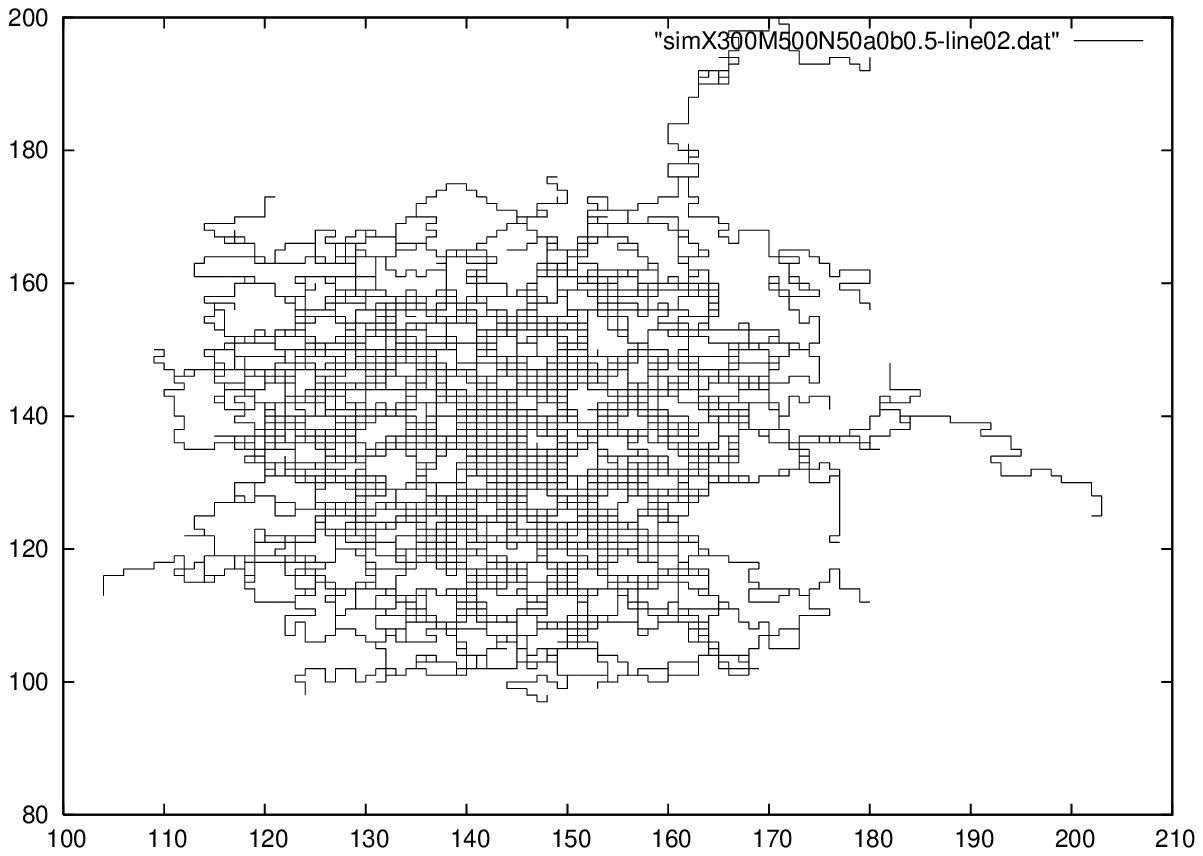}

\centerline{\bf{c)} \hspace{11em} \bf{d)} \hspace{11em}}
 \caption{Different simulated PTN
maps of 'cities' of size $300\times300$ with $M=500$ routes of
$n=50$ stations each. \bf{a)}: $a=0.0001, b=1$, \bf{b)}: $a=0.001,
b=1$, \bf{c)}: $a=0, b=0.1$, \bf{d)}: $a=0, b=0.5$.}
 \label{fig:5}
\end{figure}

The above rules resemble the preferential attachment growth rule
\cite{prefattach} favoring high-degree nodes when linking new nodes
to a network. The
principal difference of our algorithm is that at each step (2ii) we link
an {\em existing} station to a neighboring site which does not need
to be empty. New stations are then only added at the frontier of the
PTN cluster while high degree nodes (hubs) accumulate at its center.
The choice of a SAWs to model the routes may seem odd at first sight.
however, the fractal dimensions measured in PTNs \cite{Benguigui,Ferber07a}
are compatible with 2D SAW behaviour for the routes and it is obvious
that a single PTN route very seldom
intersects itself. Moreover, the
scaling properties of SAWs on disordered lattices do not change,
provided the disorder is short-range correlated \cite{saw}. In our
application this means, that even the presence of certain geographical
constraints and deviations from the square lattice
still allow for a SAW description of a PTN route.

The above described generating procedure results in a simulated
PTN of $M$ routes, each consisting of  $n$ stations. In Fig.
\ref{fig:5} we show several typical PTN configurations of
simulated cities on a $300\times 300$ square lattice with $M=500$, $n=50$ and
different values of $a$ and $b$. The parameters $a$ and $b$ allow to
discriminate between different regimes of the network evolution.
Setting $a=0$ limits every new route to start from an already existing
station. On the contrary, for $a\neq 0$, the terminal station of a new
route may be situated at any lattice site.
Therefore,  increasing $a$ allows for PTNs that consist of more
than one component (c.f. Figs. \ref{fig:5}a and \ref{fig:5}b).
The parameter $b$ on the other hand tunes the evolution of the routes:
a small value of $b$ forces the routes to propagate in parallel
({\em harnessed)} while increasing
 $b$ results in routes that cover more sites of the lattice (see
Figs. \ref{fig:5}c, \ref{fig:5}d correspondingly). In the next
section we will investigate how the numerical characteristics of the
modeled PTNs correspond to those observed for real cities.

\section{Simulation Results}
\label{sec:4}

\begin{figure}
\centering
\includegraphics[width=55mm]{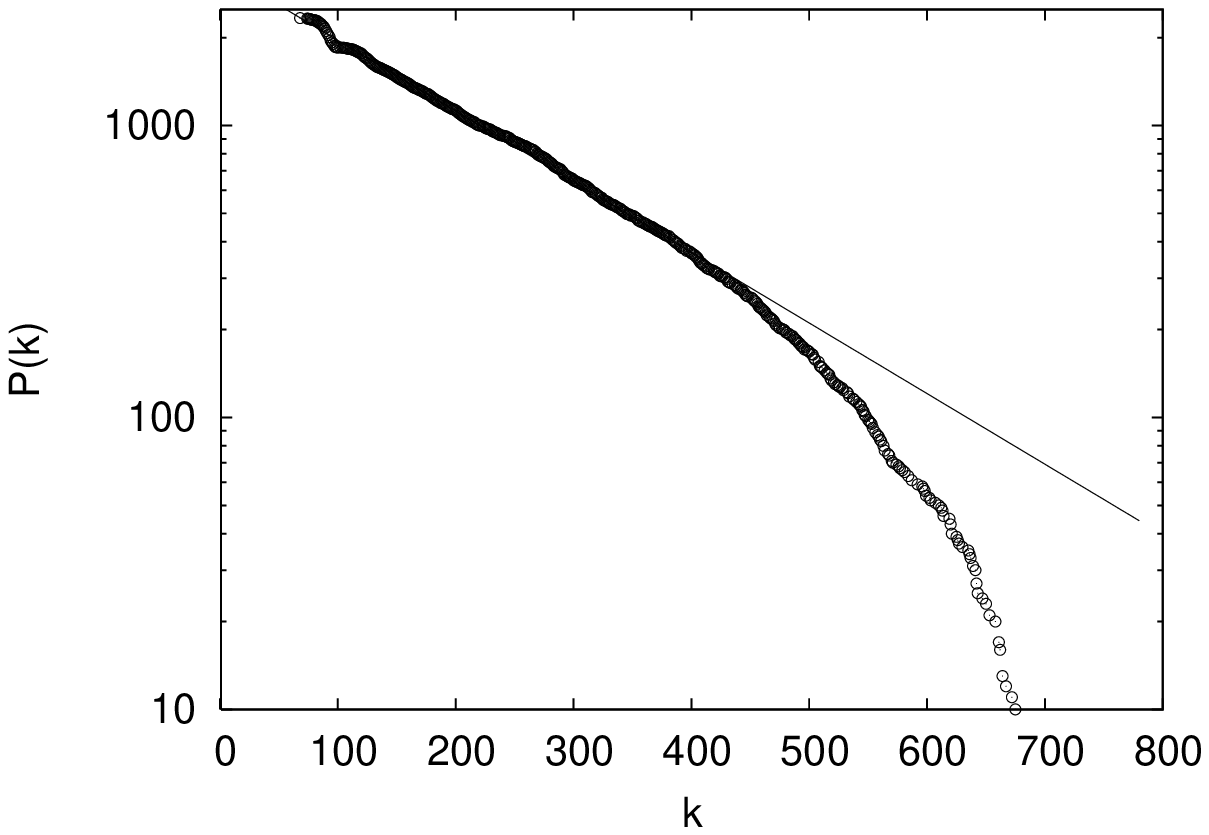}
\includegraphics[width=55mm]{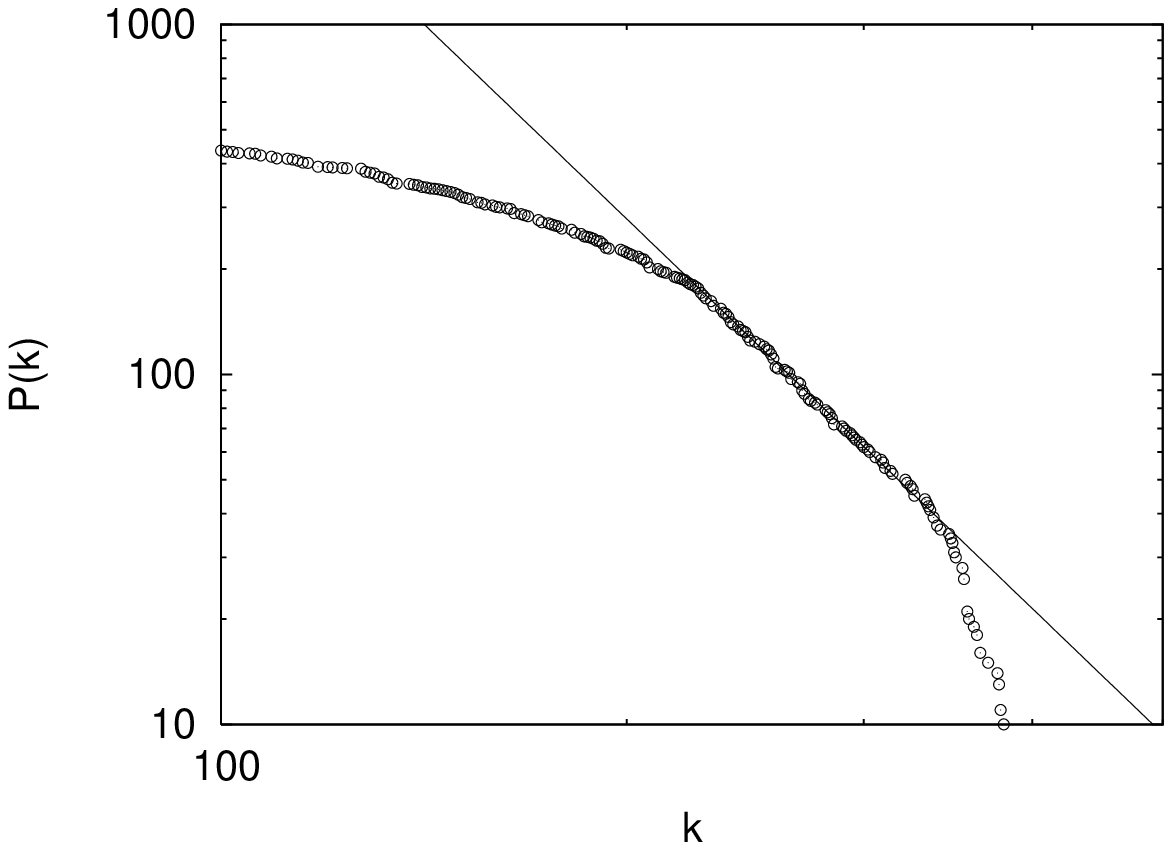}

{\bf a)} \hspace{15em} {\bf b)}
 \caption{$\mathbb{P}$-space cumulative degree distribution $P(k)$ for different simulated cities
 of $300\times 300$ sites with $M=500$ and $n=50$.
{\bf a}) $a=0$, $b=0.5$:  exponential
 in $\mathbb{P}$-space, { \bf b}) $a=0$, $b=0.1$: power law in  $\mathbb{P}$-space.}
 \label{fig:6}
\end{figure}

Postponing a more detailed analysis of the numerical simulations based
on the model of section \ref{sec:3} to a separate publication
\cite{Ferber07b}, we focus here on several principal features of
PTNs and deomstrate how they are reproduced by our model. One of them is
the behaviour of the node degree distribution $p(k)$. As it was
shown in section \ref{sec:2}, the behaviour of this function varies
for PTNs of different cities. In $\mathbb{L}$-space one often
observes $p(k)$ to be of a power-law type (\ref{eq:2})
\cite{Ferber05,Sienkiewicz05a,Xu07,Ferber07a}, however sometimes it
is governed by an exponential decay (\ref{eq:1}). Moreover, recently
 these two types of behaviour
were also observed in $\mathbb{P}$-space \cite{Ferber07a}.
A distinguished feature
of our model is that depending on the values of the evolution
parameters $a,b$ it discriminates between a power-law and an exponential
$p(k)$ in $\mathbb{P}$-space. Note that on the square lattice the
 $\mathbb{L}$-space degrees are limited to $k\leq 4$ ruling out
such an analysis.
As an
example, in Fig. \ref{fig:6} we show the cumulative node degree
distribution $P(k)$, see Eq. \ref{eq:3},
 for PTNs of two simulated cities in the $\mathbb{P}$-space.
 Changing the value of parameter
 $b$ for fixed $a$ one passes from an exponential (a straight line
 in the log-linear plot in Fig.  \ref{fig:6}a) to a power-law regime
 (a straight line in the log-log plot in Fig.  \ref{fig:6}b).

\begin{figure}
\centering
\includegraphics[width=55mm]{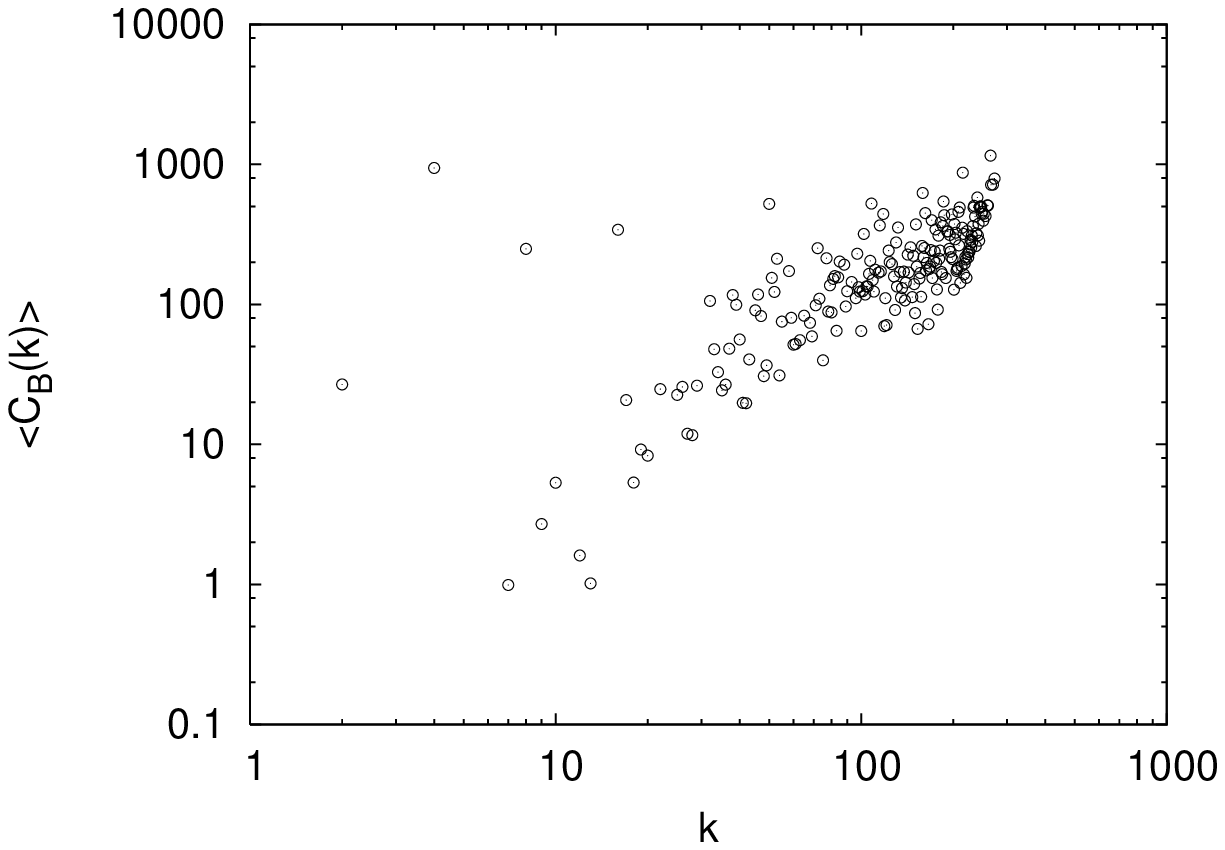}
\includegraphics[width=55mm]{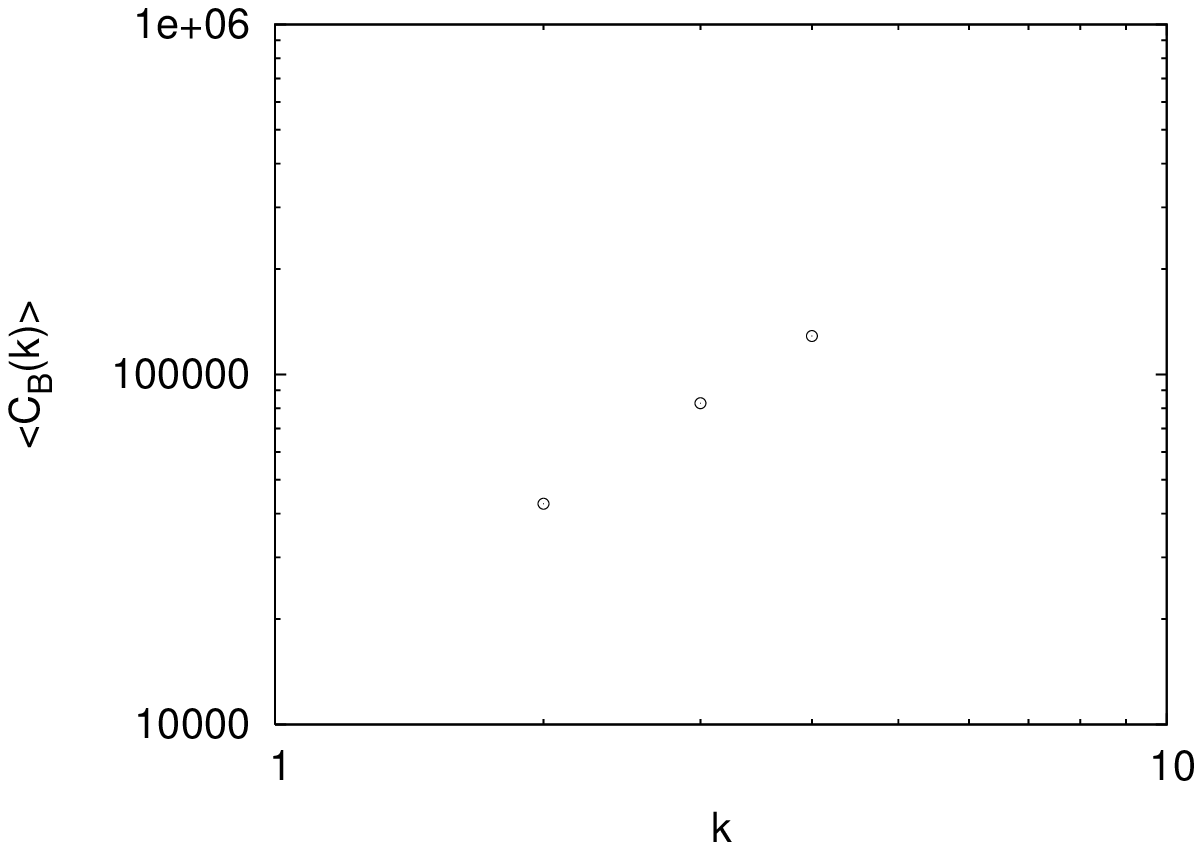}
\includegraphics[width=55mm]{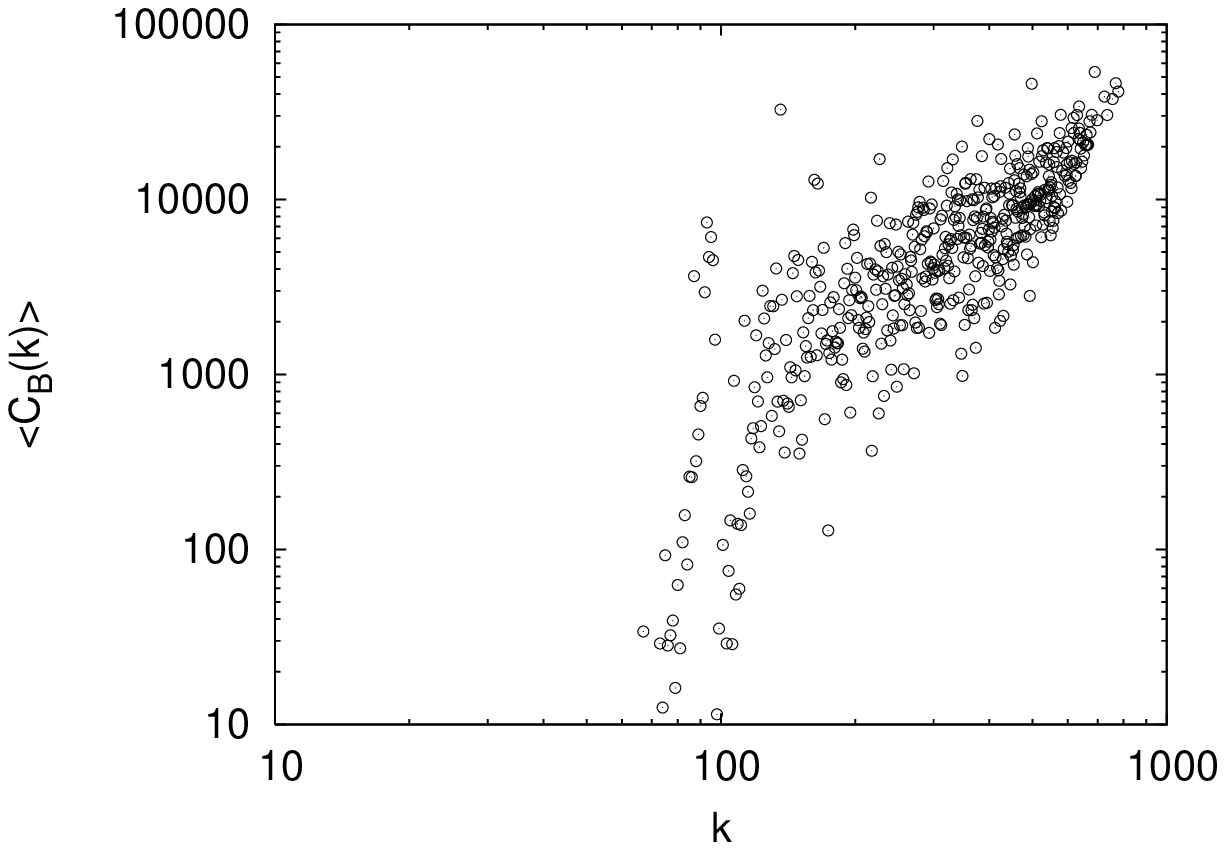}
\includegraphics[width=55mm]{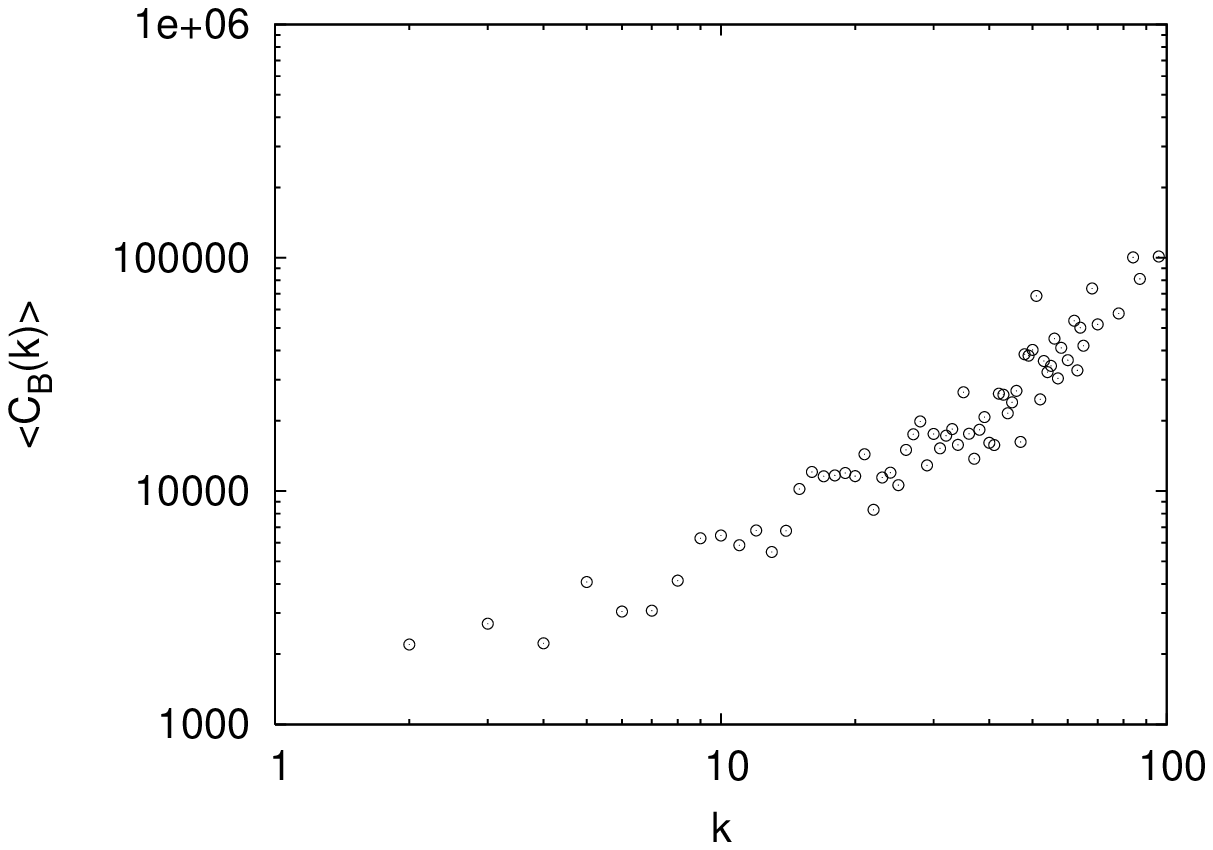}
\caption{Betweenness-degree correlation for the simulated city of
$300\times 300$ sites with $M=500$ and $n=50$ in $\mathbb{C}$-,
$\mathbb{L}$-, $\mathbb{P}$-, and $\mathbb{B}$-spaces. $a=0,
b=0.5$.}
 \label{fig:7}
\end{figure}

Another specific feature of  real world PTNs that is nicely
reproduced by our model is the harnessing effect. In Fig.
\ref{fig:3}c we show  the harness distribution $P(L,R)$ for the
PTN of a simulated city on $300\times 300$ sites with $M=500$ and
$n=50$ at $a=0$, $b=0.5$ in comparison with the same quantity for
the PTNs of Paris (Fig. \ref{fig:3}a) and of Sydney (Fig.
\ref{fig:3}b). Again, as in the case of real-world PTNs one may
speculate about power-law behaviour. Note
however, that neither for the node degree nor for the harness
distribution
we so far find a simple relation between the exponents
that may govern the scaling and the model evolution parameters
$a,b$.
Finally, let us compare the betweenness-degree
correlation. In the same way as for the PTN of Paris (Fig. \ref{fig:3}),
 we plot this function in Fig. \ref{fig:7}
for different representations for a simulated
city of $300\times 300$ sites with $M=500$ and $n=50$ for the
evolution parameters $a=0, b=0.5$. One can see an overall
qualitative agreement between the behaviour observed for the
real-world network and the simulated one (for $\mathbb{L}$ see discussion above).
In this context it is
worth to mention that the simultaneous use of different representations
(spaces) serves as a useful tool to
quantify the correspondence and differences between real word networks
as well as simulated ones.

\section{Conclusions and outlook}
\label{sec:5}

The small-world properties of public transport networks are
an everyday experience: it is easy to reach almost any
given place in a city with only a small number of changes of
transport (from table \ref{tab:1} it follows e.g. that the
connection between any two stations in Paris implies on average
$\overline{\ell_{\rm p}}-1 = 1.8$ changes), the scale-free
properties of these networks are not that evident.
Even more, scale free properties for these networks have
 sometimes been doubted. The results of our empirical analysis of PTNs of 14
major cities of the world \cite{Ferber07a,Holovatch07,Ferber07b}
together with the empirical data for several other cities
\cite{Marchiori00,Latora01,Latora02,Seaton04,Ferber05,Sienkiewicz05a,Xu07}
however
give a strong evidence of the fact that scale-free behaviour emerges
in many PTNs. Apart from our model we currently cannot
give more general arguments for this
this behaviour. Furthermore, not all PTNs seem to exhibit
power-law node degree distributions, as some
display rather an exponential decay of $p(k)$, see Table \ref{tab:1}.
Inspired by this observation we developed an evolutionary
model of self-avoiding walks on 2D lattice  which discriminates
between the above types of behaviour and recovers a number of other basic
features of the PTNs. The model applies the idea of the preferential
attachment scenario \cite{prefattach}, however with specific differences
to that standard scenario: as far as the PTNs constitute
an example of ever evolving networks, such a mechanism is not unlikely
to play a role in their growth. In more general terms, scale-free
networks have been shown in certain situation to minimize both the effort for
communication and the cost for maintaining connections
\cite{optimization,Gastner04}. Similar optimization was shown
to lead to the small world properties \cite{Mathias01} and used to
explain the appearance of power laws
\cite{zipfoptimal}. Therefore one may expect the observed scale-free behaviour
of PTNs to naturally emerge from  obvious optimization objectives
followed in their design.

One of the specific features of the PTNs we have analyzed is the harnessing
effect:  very often several routes go in parallel and pass together
several consecutive stations. While other networks with real-world
links like cables or neurons embedded in two or three dimensions
often show similar behavior, these can be studied in detail in our
present case. Our empirical analysis of the {\em harness} distribution
that quantifies this behavior indicates power-law behaviour.
 The same behaviour is inherently recovered by our model.
We found that a useful
tool to classify PTNs as well as to find correspondence between
real-world  and simulated PTNs is a comparison of the
observables in different representations (`spaces').
Furthermore, the standard network characteristics as
represented in different spaces turn out to be natural
measures for the quality of public transport in a city.

Of the many interesting further questions that are related to
this study  let us only mention the
vulnerability of PTNs to random failures and targeted attacks (see our
contribution \cite{attack} on this subject in this volume) as well as
the correlation between the topological properties
of a PTN and its geographical embedding.

\section*{Acknowledgments}
Support by the Austrian Fonds zur F\"orderung der
wi\-ssen\-schaft\-li\-chen Forschung, Project P19583 (Yu.H.) and by
the EC, Project MTKD-CT-2004-517186 (C.v.F) is gratefully acknowledged.  We thank
Dietrich Stauffer for making us aware of Ref. \cite{Benguigui}.
% BibTeX users please use
% \bibliographystyle{}
% \bibliography{}
%
% Non-BibTeX users please follow the syntax
% the syntax of "referenc.tex" for your own citations
%%%%%%%%%%%%%%%%%%%%%%%% referenc.tex %%%%%%%%%%%%%%%%%%%%%%%%%%%%%%
% sample references
% "physics"
%
% Use this file as a template for your own input.
%
%%%%%%%%%%%%%%%%%%%%%%%% Springer-Verlag %%%%%%%%%%%%%%%%%%%%%%%%%%

%
% BibTeX users please use
% \bibliographystyle{}
% \bibliography{}

%%%%%%%%%%%%%%%%%%%%%%%%%%%%%%%%%%%%%%%%%%%%%%%%%%%%%%%%%%%%%%%%%%%%%%  }

%%%%%%%%%%%%%%%%%%%%%%%%%%%%%%%%%%%%%%%%%%%%%%%%%%%%%%%%%%%%%%%%%%%%%%

\printindex
\end{document}